\documentclass[lettersize,journal]{IEEEtran}
\usepackage{mathtools}
\usepackage{enumitem}
\usepackage{multirow}
\usepackage{amsthm}
\usepackage{amssymb}
\usepackage{amsmath}
\usepackage{blindtext}
\usepackage{graphicx}
\usepackage{algpseudocode}
\usepackage{algorithm}
\usepackage{booktabs}
\usepackage{multicol}
\usepackage[font=scriptsize]{caption}
\usepackage{verbatim}
\usepackage{float}

\usepackage{xcolor}

\theoremstyle{plain}% Theorem-like structures provided by amsthm.sty
\newtheorem{theorem}{Theorem}[section]

\newtheorem{corollary}[theorem]{Corollary}

\theoremstyle{definition}

\theoremstyle{remark}

\begin{document}

\title{Distributed Marginal Return: A Communication-Efficient Algorithm for Multi-Agent Weapon Target Assignment}

\author{Kamal Mammadov, Glen Pearce, Helen Krause, Damith C. Ranasinghe
\thanks{D. Ranasinghe is with the School of Computer Science and Information Technology, Adelaide University, Australia.}
\thanks{G. Pearce \& H. Krause is with the Defence Science and Technology Group, Australia.}
}

\maketitle

\begin{abstract}
The Weapon-Target Assignment (WTA) problem in distributed multi-agent systems is challenged by limited situational awareness and communication constraints. Existing distributed methods, such as auction-consensus algorithms, typically assume task-value independence, a simplification that fails to capture the coupled effects of multi-weapon assignments. This paper introduces the Distributed Marginal Return (DMR) algorithm, a novel decentralized approach that generalizes auction-consensus frameworks. DMR enables agents to compute bids based on marginal return and uses a consensus protocol to resolve conflicts, allowing multiple non-conflicting proposals to be accepted in parallel for cooperative optimization with minimal communication. A key theoretical contribution is our proof that the marginal returns of non-conflicting bids are additive, a property that ensures global consistency and validates our parallel acceptance mechanism. Extensive simulations demonstrate that DMR significantly outperforms baseline algorithms in both solution quality and communication efficiency, offering a scalable and robust solution for real-world distributed command and control applications.
\end{abstract}


\section{Introduction}
The Weapon-Target Assignment (WTA) is a fundamental problem in military decision-making, involving the optimal allocation of defensive or offensive resources to neutralize adversarial threats. Its applications span missile defense, unmanned combat systems, and multi-stage battlefield scenarios, where rapid and efficient resource allocation directly impacts mission success \cite{survey2024, survey2025, Sam_Matlin1970}. While centralized algorithms have historically dominated WTA research, the emergence of distributed multi-agent systems, with constraints on communication, situational awareness, and operational security, demands novel decentralized approaches. 

While the WTA problem is inherently rooted in military operations, it is crucial to recognize its broader significance from a defensive and peacekeeping perspective. Modern applications of WTA, particularly within distributed missile defence \cite{Yuanzhi2022} and autonomous interception systems \cite{ZhiLin2025}, are fundamentally designed to neutralize incoming threats and protect vulnerable populations. Furthermore, by optimizing resource allocation and maximizing operational precision, advanced WTA algorithms minimize collateral damage and help prevent the inadvertent escalation of conflicts. In this context, the development of robust, decentralized defensive capabilities acts as a vital deterrent \cite{Zimo2026}, ultimately contributing to the preservation of global peace and stability.

The Weapon-Target Assignment (WTA) problem has evolved significantly since its inception, shaped by diverse objectives and computational challenges. The foundational work by \cite{Manne1958} framed WTA as an optimization task to minimize the expected survival value of enemy targets, establishing a mathematical basis for subsequent research. \cite{Soland1973} streamlined this framework by excluding weapon effectiveness probabilities, simplifying computational demands and enabling early algorithmic exploration. Over time, model complexity escalated as researchers incorporated multifaceted objectives. \cite{Murphey2000} prioritized maximal target destruction, \cite{Lee2002} emphasized global damage probability, \cite{Paradis2005} focused on safeguarding defensive assets, and \cite{Song2009} optimized cost efficiency in weapon deployment. \cite{lloyd1986} established the weapon-target assignment problem is NP-complete.

Early WTA formulations focused on static scenarios, where target valuations, weapon effectiveness, and battlefield parameters remain constant. Seminal works like \cite{Kwon1999640} minimized weapon assignment costs via integer programming with linear objectives and nonlinear survivability constraints, while \cite{Shang2008221} incorporated protected asset values and probabilistic interdependencies to optimize for minimal expected damage. Subsequent extensions introduced practical constraints, \cite{grouping_constraint} enforced weapon allocation bounds per target, \cite{timing_cost_function} penalized deviations from desired weapon arrival times, and \cite{artillery_attack} modelled time-decaying destruction probabilities for fire scheduling. For unmanned systems, \cite{ucav_routing_problem, uav_defence_technology} combined WTA with vehicle routing, optimizing unmanned combat aerial vehicle (UCAV) paths, weapon configurations, and attack windows. This evolution led to a natural division between static WTA (fixed targets/weapons, time-agnostic) and dynamic WTA, which incorporates temporal adaptability as battlefield parameters, such as threat levels, target positions, and enemy tactics, evolve rapidly.

Dynamic WTA formulations address this temporal complexity through two primary approaches, decomposing the problem into sequential static (m-stage) phases or employing Shoot-Look-Shoot (SLS) methods for fully dynamic, non-segmentable scenarios. Building on these distinctions, dynamic frameworks continuously update weapon-target pairings across engagement rounds. For instance, \cite{Xin2011598} defined multi-stage problems where weapon-target probabilities and asset values evolve over time, requiring sequential optimization under uncertainty. Others, like \cite{Murphey1999406}, framed dynamic WTA as stochastic programs to balance immediate assignments against future risks. The trilevel ``surface-to-air missile defense game'' by \cite{Boardman2017417} extended WTA into adversarial game theory, optimizing SAM placements against anticipated attacker strategies. 

Exact methods, such as integer programming and branch-and-bound algorithms \cite{Kline20171}, emerged to guarantee global optimality for simplified or small-scale instances. However, their exponential time complexity renders them impractical for real-world scenarios involving large-scale, multi-objective allocations. To address these limitations, heuristic approaches gained prominence. The Maximum Marginal Return (MMR) algorithm \cite{Denbroeder1959322} exemplified this shift, employing a greedy iterative process to assign weapons based on incremental utility gains, thereby achieving near-optimal solutions with polynomial-time scalability. An algorithm based on quantum computing \cite{quantum_alg} was also proposed to handle the exponential complexity.

For larger-scale problems, meta-heuristics have dominated. Genetic Algorithms (GA) \cite{Metler1990, Chen20092006} evolve populations of assignment strategies through mechanisms such as crossover and mutation, whereas Ant Colony Optimization \cite{Lee2002287} and Particle Swarm Optimization \cite{Johansson2010} mimic biological swarming behaviours. Simulated annealing \cite{Tokgöz201339} and Tabu Search \cite{Ahuja20071136} employ stochastic neighbourhood searches, escaping local optima through probabilistic acceptance rules. Recent efforts, such as \cite{Andersen2022}, have linearized nonlinear survivability constraints through probability chains, thereby enabling mixed-integer programming approximations.

While centralized optimization has historically dominated WTA research, its practical application is often limited. Modern military operations are characterized by distributed assets, such as swarms of unmanned aerial vehicles (UAVs), networked ground sensors, and coordinated missile defence batteries; where a central command node creates a single point of failure and a communication bottleneck. Indeed, nearly all contemporary applications of WTA are inherently distributed. Furthermore, a centralized architecture does not eliminate the need for communication; agents must still transmit situational awareness data to the central node, a process that consumes bandwidth and introduces latency. Decentralized approaches, in contrast, embrace this distributed reality by enabling agents to coordinate directly, enhancing robustness, scalability, and adaptability in dynamic, communication-constrained environments. The transition to distributed WTA, however, introduces unique challenges, including consensus formation and decentralized optimization, which necessitate novel algorithmic solutions.

Our work focuses on addressing these challenges by developing a distributed WTA framework that enables cooperative decision-making among autonomous agents. The following section considers related methods for distributed multi-agent assignment problems.
\subsection{Related Work}
Advancements in decentralized task allocation have spurred the development of distributed algorithms that address complex multi-agent assignment problems, including weapon-target assignment. A seminal contribution in this area is the Consensus-Based Auction Algorithm (CBAA) introduced in \cite{CBAA}. CBAA is designed for single-assignment scenarios and operates through iterative Auction and Consensus phases. In the auction phase, each agent asynchronously places bids on available tasks, generating local evaluations based on task-specific reward functions. The subsequent consensus phase enables agents to share and update their winning bid information with neighbouring agents, ensuring that the network converges on a globally consistent assignment despite the absence of a centralized decision node. 

An extension of CBAA, known as the Consensus-Based Bundle Algorithm (CBBA), was developed to tackle more complex multi-assignment problems where agents can be allocated bundles of tasks, and the order in which the tasks are completed by each agent effects the reward function. CBBA introduces additional procedures, such as bundle construction and advanced conflict-resolution mechanisms. However, when each agent is limited to a single assignment, as is the case with weapon-target engagement, CBBA simplifies exactly to the CBAA algorithm. In this single-assignment setting, CBAA is effectively a special case of CBBA, making it ideally suited for our application. Consequently, our work builds directly on the CBAA framework, leveraging its iterative auction-consensus structure to develop a Distributed Marginal Return algorithm that meets the operational constraint of one weapon per target.

Other works in distributed agent task allocation include \cite{GRAPE}, which formulates the multi-robot task allocation problem as a hedonic game where each agent’s utility depends not only on the task it performs but also on the number of collaborators involved. In this framework, agents are assumed to be single-task robots, and the overall goal is to maximize the global utility by partitioning agents into coalitions assigned to tasks (including a “void” task for those opting out). A key innovation in GRAPE is the introduction of SPAO (Single-Peaked-At-One) preferences, which encapsulate the idea of social inhibition, agents naturally prefer to work with fewer collaborators since their individual performance diminishes as more agents join the same task. This preference structure ensures that a Nash stable partition exists under certain conditions, meaning that no agent has an incentive to unilaterally deviate from its current assignment. The paper rigorously proves both the existence of such a stable partition and provides bounds on the number of iterations required for convergence, specifically showing that a Nash stable partition is reached in polynomial time.

In addition to its theoretical contributions, the GRAPE paper proposes a fully decentralized algorithm that enables agents to reach a Nash stable partition without centralized coordination. Each agent iteratively evaluates its local preferences over task-coalition pairs and makes decisions based solely on locally available information and limited communication with neighbouring agents. A crucial component of this algorithm is a distributed mutex subroutine, which resolves conflicts among concurrently updated partitions and ensures convergence even in asynchronous settings. Through this mechanism, the algorithm guarantees that all agents eventually agree on the same partition, thus aligning their individual decisions with a globally stable outcome. The GRAPE approach, with its game-theoretic foundation and robust convergence properties, offers valuable insights into the design of distributed task allocation algorithms, complementing auction-based methods by addressing the interplay between individual preferences and collective stability.

Later, \cite{BidGroup} extended the work \cite{CBAA} to G-CBBA, offering a systematic approach to grouping agents based on task preferences as represented by their initial bids. Additionally, \cite{RobotTaskAllocation} proposed a heuristic for multi-robot task allocation in warehouse settings, aiming to minimize the total path length traveled by robots while satisfying capacity constraints. Moreover, \cite{One-to-many} developed the One-to-Many Bipartite Matching Algorithm, which forms coalitions by allocating robots to tasks in a cost-minimizing manner using a variant of classical bipartite matching. Finally, \cite{MarketAlg} examined the static Law Enforcement Problem, in which patrols are allocated to police units in a city while accounting for travel times, task importance, and workload constraints. They proposed a task allocation algorithm that achieves both fairness (in terms of envy-free allocations) and efficiency (Pareto optimality) through a Fisher market framework.

\vspace{2mm}
%\noindent\textbf{Summary.~}In summary, while the literature on Weapon-Target Assignment is extensive, a critical gap remains in its application to distributed multi-agent systems. Centralized methods are ill-suited for modern, decentralized battlefields, which are defined by communication constraints and the need for operational resilience. Concurrently, prominent distributed assignment algorithms like the Consensus-Based Auction Algorithm (CBAA) are fundamentally designed for one-to-one task allocation. They cannot effectively handle the core complexity of WTA: evaluating the coupled, non-linear returns from assigning multiple weapons to a single target. This limitation necessitates a new class of algorithm capable of achieving consensus on complex, interdependent assignments with high communication efficiency. Our work directly addresses this gap.

\subsection{Our Contributions}

While the contemporary literature contains numerous recent advancements in automated target assignment algorithms, several prominent state-of-the-art methods, such as \cite{Hendrickson2021}, address fundamentally divergent problem formulations or impose highly restrictive network constraints that completely preclude direct methodological comparison with the Distributed Marginal Return (DMR) framework. For instance, the Marginal-Return-Based Constructive Heuristic (MRBCH) proposed by \cite{Xin2019} exploits marginal return logic highly efficiently, but it is formulated specifically and exclusively for the joint Sensor-Weapon-Target Assignment (S-WTA) problem. MRBCH fundamentally relies upon a rigid centralized architecture, utilizing a three-dimensional auxiliary decision matrix that explicitly requires instantaneous, global state knowledge of all interconnected resources to execute its sequential lookup operations. Because MRBCH rigidly assumes a zero-latency, fully connected command topology, attempting to benchmark it against DMR's communication-constrained, fully decentralized operational architecture would yield an inherently incongruous comparison of both communication overhead and dynamic scalability.

Similarly, highly popular consensus-based market approaches, such as the Multi-Target Consensus-Based Auction Algorithm (MTCBAA) introduced by \cite{aerospace2022}, successfully distribute computational loads but introduce distinct, prohibitive algorithmic constraints. MTCBAA is tailored specifically for Beyond-Visual-Range (BVR) combat environments, heavily utilizing complex geometric advantage functions rather than standard probabilistic survival metrics. More critically, MTCBAA absolutely relies on a synchronous consensus phase to resolve greedy bidding conflicts across the swarm, mathematically altering the theoretical optimization bounds to a guaranteed 50 percent minimum performance threshold. The strict requirement for synchronous communication rounds renders MTCBAA highly sensitive to the ad-hoc, intermittent, and chaotic network topologies that DMR is explicitly designed to navigate natively, making direct quantitative comparisons of convergence rates under degraded communication environments methodologically inequitable.

Consequently, the later comparative evaluations in this manuscript focus solely on \cite{CBAA} and \cite{GRAPE}, as these were the only works whose problem formulations aligned sufficiently with ours to justify a direct, one-to-one comparison, thereby ensuring that all baseline performance comparisons remain methodologically sound.

\section{Problem Formulation}\label{section_problem_formulation}
Consider a distributed multi-agent weapon-target assignment (WTA) problem defined over a network of autonomous agents $\mathcal{A}$ (e.g., platforms or units) and a set of targets $\mathcal{T}$. Unlike classical centralized WTA models, our formulation explicitly incorporates distributed constraints governing communication and situational awareness. 

The communication topology of the multi-agent system is modelled as an undirected graph $\mathcal{G} = (\mathcal{A}, \mathcal{E})$, where $\mathcal{E} \subseteq \mathcal{A} \times \mathcal{A}$ represents the set of available communication links. An agent $i \in \mathcal{A}$ can only exchange information directly with its spatial neighbours, defined by the set $\mathcal{N}_i = \{l \in \mathcal{A} \mid (i,l) \in \mathcal{E}\}$.

Furthermore, each agent operates under partial observability. Rather than having global situational awareness, an agent $i$ only observes a local subset of the targets, denoted as $\mathcal{T}_i \subseteq \mathcal{T}$. Each agent $i$ possesses a local set of weapons $\mathcal{W}_i$. The local information state available to agent $i$ is strictly limited to its own weapon capabilities, deployment costs, and its observed targets. The true global state remains unknown to any single agent.

Let $x_{ikj} \in \{0,1\}$ denote a binary decision variable equal to 1 if agent $i$ assigns weapon $k \in \mathcal{W}_i$ to target $j \in \mathcal{T}$, and 0 otherwise. To seamlessly bridge the local observability constraint with the global assignment problem, we define the probability of destruction $p_{ikj} \in [0,1]$ for weapon $k$ from agent $i$ against target $j$. If a target is outside the observable and engageable range of agent $i$ (i.e., $j \notin \mathcal{T}_i$), we naturally enforce $p_{ikj} = 0$. Each target $j \in \mathcal{T}$ is associated with a priority weight $w_j \geq 0$, representing its strategic importance, and $c_{ik} \geq 0$ denotes the local cost of deploying weapon $k$ from agent $i$.

The global optimization problem is subject to two primary constraints. First, each weapon can be assigned to at most one target, formalized as:
\begin{equation}\label{constraint1}
\sum_{j \in \mathcal{T}} x_{ikj} \leq 1 \quad \forall i \in \mathcal{A},\ \forall k \in \mathcal{W}_i.
\end{equation}
Second, the decision variables are restricted to binary values to enforce discrete assignments:
\begin{equation}\label{constraint2}
x_{ikj} \in \{0,1\} \quad \forall i \in \mathcal{A},\ \forall k \in \mathcal{W}_i,\ \forall j \in \mathcal{T}.
\end{equation}

The objective function balances two competing goals: minimizing the total surviving target value and minimizing weapon utilization costs. The total surviving target value, denoted $T(\mathbf{x})$, aggregates the expected residual value of all targets after weapon assignments. For each target $j$, its survival probability $P_j^\text{survive}$ is computed as the product of probabilities that no assigned weapon destroys it:
\begin{equation}\label{P_j}
P_j^\text{survive} = \prod_{i \in \mathcal{A}} \prod_{k \in \mathcal{W}_i} \left(1 - p_{ikj}\right)^{x_{ikj}}.
\end{equation}
The total surviving target value is then given by:
\begin{equation}\label{T_x}
T(\mathbf{x}) = \sum_{j \in \mathcal{T}} w_j P_j^\text{survive}.
\end{equation}
The weapon utilization cost $C(\mathbf{x})$ sums the deployment costs of all assigned weapons:
\begin{equation}\label{C_x}
C(\mathbf{x}) = \sum_{i \in \mathcal{A}} \sum_{k \in \mathcal{W}_i} c_{ik} \sum_{j \in \mathcal{T}} x_{ikj}.
\end{equation}
The combined objective function $J(\mathbf{x})$ seeks to minimize the weighted sum of these two components:
\begin{equation}\label{J_x}
J(\mathbf{x}) = T(\mathbf{x}) + C(\mathbf{x}),
\end{equation}
and the global optimization problem is formulated as:
\begin{equation}\label{min_J_x}
\min_{\mathbf{x}} \ J(\mathbf{x}),
\end{equation}
where $\mathbf{x}$ denotes the vector of all decision variables $x_{ikj}$.

This formulation defines the global objective that the network of agents must cooperatively minimize. Because no central node exists and no individual agent possesses the full global state or the complete decision vector $\mathbf{x}$, the system must rely on distributed algorithms operating over the communication graph $\mathcal{G}$ to evaluate assignments and reach consensus on a globally consistent allocation based purely on local information.

\begin{figure}[h]
\centering
\includegraphics[width=1.0\columnwidth]{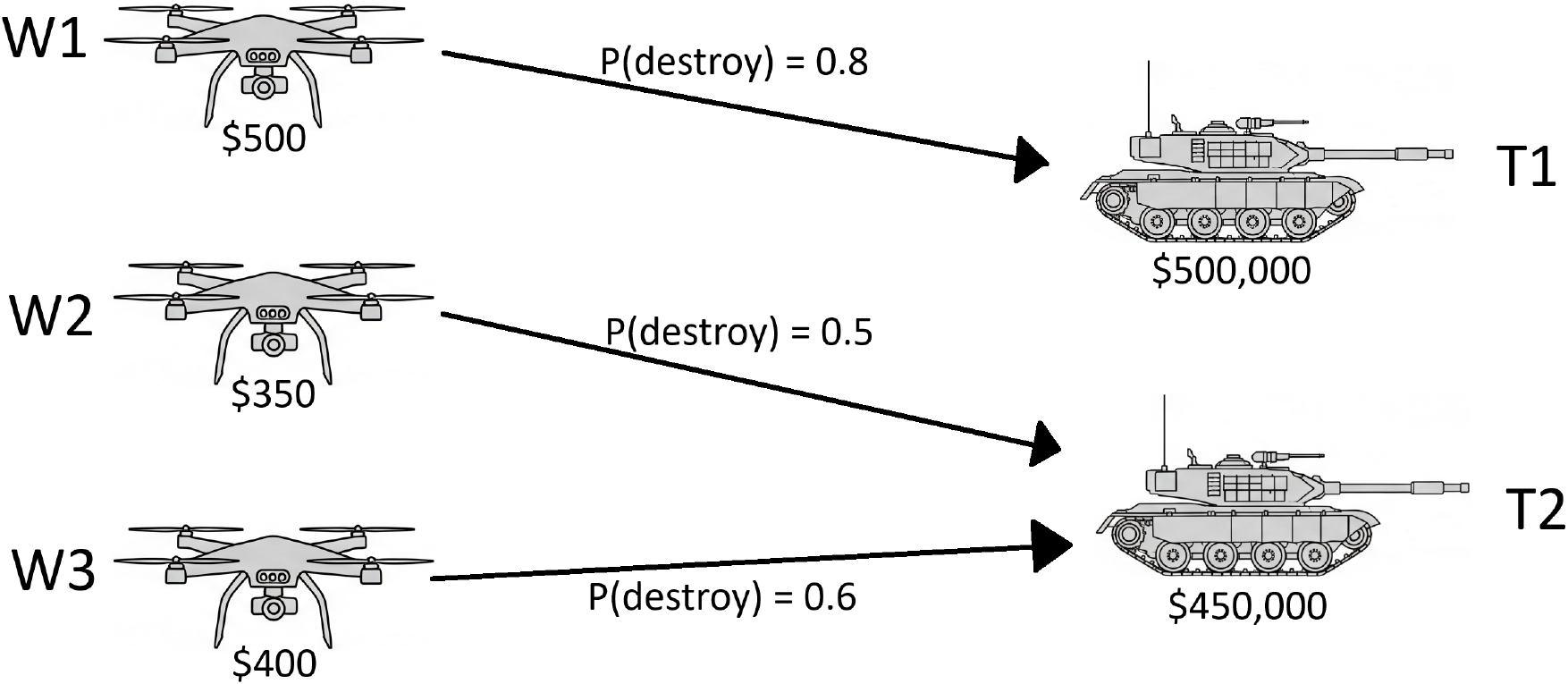}
\caption{An illustrative example of a weapon-target assignment problem. Three weapons (W1,W2,W3) with varying costs must be assigned to two high-value targets (T1,T2). The objective is to minimize the total surviving target value while accounting for the cost of the weapons deployed.}
\label{fig:wta_example}
\end{figure}
To provide a tangible context for our formulation, consider the example scenario depicted in Figure \ref{fig:wta_example}. Here, a set of autonomous weapons must be allocated to neutralize a set of high-value targets. Each weapon has an associated deployment cost $c_{ik}$, and each target has a strategic value $w_j$, which the assignment seeks to minimize. This simple example encapsulates the core trade-off in the WTA problem: achieving maximum mission impact (destroying valuable targets) while minimizing resource expenditure.

\section{Distributed Marginal Return Algorithm}\label{section_dmr}

To intuitively conceptualize the proposed architecture, the end-to-end system flow operates as follows. The system inputs consist of decentralized, local information: each agent's partial situational awareness, target priority weights, and localized weapon capabilities (specifically, individual kill probabilities and utilization costs). These inputs are processed through a sequential, distributed mechanism comprising an iterative loop of Auction and Consensus phases. During this processing stage, agents independently compute assignment bids based on marginal returns and subsequently utilize a multi-hop communication relay to collectively resolve conflicting bids across the network. The definitive output of this cooperative system is a globally consistent, binary weapon-target assignment matrix $\mathbf{x}$ that minimizes the overarching objective function, achieved entirely without a central command node.

%\subsection{Motivation}
Inspired by the CBAA algorithm, our goal is to develop a fully distributed approach that incorporates two key phases: an auction phase, where each agent independently proposes deviations from the current assignment, and a consensus phase, where agents communicate to collectively agree on the final allocation. The design of this algorithm focuses on addressing several critical considerations:
\begin{itemize} 
    \item Each agent has knowledge solely of the characteristics of the weapons it possesses, specifically, the weapon utilization costs and probabilities of kill for each target. Agents do not have access to information about other agents in the battlespace and can only propose bids based on their own weapon set.
    \item The only information shared globally across the battlespace pertains to the targets. This includes a unique target ID and an associated target value for each target. Beyond this, all other details, such as the weapon characteristics of other agents and even the total number of agents in the battlespace, are assumed to remain unknown to each agent. 
    \item To enhance operational security, the distributed algorithm should be designed to minimize inter-agent communication, as electromagnetic emissions can reveal the positions of agents, making them susceptible to detection or localization by adversaries. 
\end{itemize}
The key manner which will differentiate our approach with that of CBAA, lies in the specific design of our algorithm to address the weapon-target assignment problem. Unlike CBAA, our algorithm allows for assigning multiple weapons to a single target, to maximize the probability of kill in scenarios where a single weapon is insufficient to achieve the desired outcome.

A defining feature of the CBAA algorithm is its distributed "divide-and-conquer" approach to assigning agents to tasks. In CBAA, each agent maintains two vectors of length $N_t$, where $N_t$ represents the number of tasks. During the auction phase, each agent asynchronously bids on a task if its task cost it less than the current bid. In the subsequent consensus phase, the winning bid for each task is evaluated independently. This approach effectively decomposes the agent-task allocation problem into $N_t$ separate and independent optimization problems, significantly simplifying the overall complexity. This aspect is crucial as it enables multiple bids to be accepted within a single consensus phase, allowing for substantial adjustments to the current assignment in each iteration. As a result, the algorithm can converge to an effective solution with minimal inter-agent communication, enhancing both efficiency and operational security.

A natural question arises, can the CBAA algorithm be generalized or extended to address the weapon-target assignment problem? Specifically, can each target in the weapon-target assignment problem be treated as a separate optimization problem, with weapons assigned to each target independently, similar to how tasks are handled in CBAA? The initial answer is no. Unlike the agent-task allocation problem, the weapon-target assignment problem allows multiple weapons to be assigned to a single target. This introduces a critical interdependence, the impact of assigning or reassigning a single weapon cannot be evaluated in isolation, as the reduction in the total assignment cost depends on the combined effect of other proposed deviations. In other words, the potential improvement from a single weapon reassignment is inherently coupled with the outcomes of other accepted deviations.

A potential compromise might involve accepting bids from only a single agent during each consensus phase, as is the case for the GRAPE algorithm \cite{GRAPE}. However, such an approach would require numerous consensus phases to converge to a good solution, necessitating extensive rounds of inter-agent communication. To address this limitation, this manuscript proposes a more sophisticated mechanism that enables the acceptance of multiple bids from multiple agents within each consensus phase, by employing a conflict-resolution process to eliminate conflicting bids. The proposed algorithm leverages the concept of marginal return, which evaluates the reduction in total assignment cost if a proposed deviation from the current assignment is accepted. We argue that this Distributed Marginal Return (DMR) approach serves as a natural generalization of CBAA, tailored specifically to the complexities of the weapon-target assignment problem.

\subsection{Description}
The Distributed Marginal Return (DMR) algorithm, like CBAA, operates in two key phases: the Auction phase and the Consensus phase. During the Auction phase, each agent independently evaluates how it can improve the total assignment cost of the current allocation by considering only its own weapon set. Based on this evaluation, the agent generates a list of bids that represent potential improvements. In the Consensus phase, agents communicate their bids with one another and collectively accept multiple conflict-free bids according to the criteria defined below. The DMR algorithm is iterative and can be executed over any number of rounds, with each round comprising an Auction and Consensus phase. Over successive rounds, the algorithm converges to a local minimum, where no single weapon reassignment from any agent can further reduce the assignment cost, resulting in all agents having empty bid lists. However, to reduce inter-agent communication, users may choose to terminate the algorithm before convergence is reached.

\subsubsection{Auction Phase}
At the beginning of each round \( t \), every agent \( i \in \mathcal{A} \) locally maintains the current assignment in a data structure denoted as \( \mathbf{x}_i(t) \), along with the corresponding objective value \( J(\mathbf{x}_i(t)) = T(\mathbf{x}_i(t)) + C(\mathbf{x}_i(t)) \). The data structure \( \mathbf{x}_i(t) \) represents the assignments accepted by all agents across the battlespace at round \( t \). Initially, at \( t = 0 \), no weapons are assigned, so \( \mathbf{x}_i(0) = \mathbf{0} \), resulting in an initial objective value of \( J(\mathbf{x}_i(0)) = \sum_{j \in \mathcal{T}} w_j \), where \( w_j \) denotes the value of each target \( j \).

Following the initial setup, the Auction phase of the Distributed Marginal Return (DMR) algorithm involves each agent systematically evaluating and prioritizing potential weapon assignments to optimize the overall objective function \( J(\mathbf{x}_i(t)) \). At each round \( t \), agent \( i \) focuses exclusively on its own weapon set \( \mathcal{W}_i \), generating a comprehensive list of possible single-weapon assignments or reassignments. This list is organized into a queue where each entry corresponds to a specific weapon and its potential target assignment. The proposals within this queue are sorted in descending order based on their \textit{marginal return}, which quantifies the expected reduction in the total assignment cost \( J(\mathbf{x}_i(t)) \) if the proposed assignment is implemented. Only proposals that result in a positive marginal return are included in the queue, ensuring that every bid contributes to the reduction in total assignment cost.

Each element in the queue is encapsulated in a structured data format comprising six key components: the \textit{Weapon ID}, the \textit{Designated Target ID}, the \textit{Previous Target ID}, \textit{Probability of kill}, \textit{Weapon utilization cost}, and the \textit{Marginal Return} value. The \textit{Weapon ID} identifies the specific weapon from \( \mathcal{W}_i \) that is being proposed for assignment or reassignment. The \textit{Designated Target ID} specifies the target that the weapon is intended to engage; if the agent opts to drop the weapon, this field remains empty. The \textit{Previous Target ID} indicates the previous target assignment of the weapon, if any, allowing the agent to track changes in weapon allocation. The next two fields denote the parameters of this proposed weapon-target allocation. Finally, the \textit{Marginal Return} represents the quantified benefit in terms of cost reduction that would result from approving the proposed assignment. 

\subsubsection{Consensus Phase}
During the Consensus phase, each agent \( i \in \mathcal{A} \) disseminates its ranked queue of single-weapon assignments, generated during the Auction phase, to all other agents within the battlespace. Upon receiving the queues from its neighbours, agent \( i \) amalgamates these inputs, including its own queue, into a unified aggregate queue. This combined queue is then sorted in descending order based on the marginal return values, ensuring that assignments offering the greatest reduction in the objective function \( J(\mathbf{x}_i(t)) \) are prioritized.

To accommodate constrained communication environments, agents operate in a decentralized relay mode, continuously forwarding received bid queues to their available spatial neighbours $\mathcal{N}_i$. To maintain strict global consistency and prevent the propagation of asynchronous conflicts, the consensus phase at each agent is synchronized to commence only once the relayed queues from all participating agents have been successfully aggregated.

Once the aggregate queue is established, each agent \( i \) proceeds to update the global weapon-target assignment from \( \mathbf{x}_i(t) \) to \( \mathbf{x}_i(t+1) \) by sequentially processing the elements of the queue. The update procedure follows a two-step algorithm while the queue is not empty:
\begin{enumerate}
    \item \textbf{Implementation of assignments:} The agent implements the single-weapon assignment corresponding to the first element of the aggregate queue, thereby incorporating the most beneficial deviation into the current assignment.
    \item \textbf{Conflict resolution:} Subsequently, the agent deletes the \textbf{implemented assignment} and all \textbf{conflicting} assignments from the queue.
\end{enumerate}

A \textbf{proposed assignment} (any other element in the queue) is deemed \textbf{conflicting} if it satisfies either of the following conditions:
\begin{itemize}
    \item[\textbf{(C1)}] The \textbf{proposed assignment} utilizes the same weapon from the same agent as the \textbf{implemented assignment}. This restriction ensures that each weapon is allocated to at most one target, adhering to physical constraints.
    \item[\textbf{(C2)}] In the \textbf{proposed assignment}, the \textit{Designated Target ID} or \textit{Previous Target ID} matches any of the \textit{Designated Target ID} or \textit{Previous Target ID} of the \textbf{implemented assignment}. This condition is critical because the marginal return calculations for these assignments depend on specific assumptions regarding the current state of target engagements. If these assumptions are violated, the previously computed marginal returns become invalid, necessitating the exclusion of these conflicting assignments.\footnote{Notably, if any of the Target ID fields are empty, for the purposes of the DMR algorithm, we do not consider two empty fields to match.}
\end{itemize}

By enforcing these conflict resolution criteria, the Consensus phase ensures that the global weapon-target assignment remains coherent and that the integrity of the marginal return evaluations is preserved. The subsequent theorem provides a formal theoretical foundation for this property. In the following, we define a \textit{single-weapon deviation} as a hypothetical modification to the assignment matrix where only the allocation of a single weapon is changed from the current consensus.

\begin{theorem}\label{Non-conflicting_Theorem}
    Let \(\mathbf{x}\) denote the current global consensus on weapon-target allocations, and consider two proposed single-weapon assignments \(\Delta \mathbf{x}_1\) and \(\Delta \mathbf{x}_2\) that deviate from \(\mathbf{x}\). Define their marginal returns as:
    \begin{align*}
        \textbf{MR}(\Delta \mathbf{x}_1) &\coloneqq J(\mathbf{x}) - J(\mathbf{x}+\Delta \mathbf{x}_1),\\
        \textbf{MR}(\Delta \mathbf{x}_2) &\coloneqq J(\mathbf{x}) - J(\mathbf{x}+\Delta \mathbf{x}_2).
    \end{align*}
    Additionally, let
    \[
    \textbf{MR}(\Delta \mathbf{x}_1 + \Delta \mathbf{x}_2) \coloneqq J(\mathbf{x}) - J(\mathbf{x}+\Delta \mathbf{x}_1 + \Delta \mathbf{x}_2)
    \]
    denote the combined effect of implementing both of the \textbf{proposed assignments}. If \(\Delta \mathbf{x}_1\) and \(\Delta \mathbf{x}_2\) are \textbf{non-conflicting} (i.e., they do not satisfy conditions (C1) or (C2)), then:
    \[
    \textbf{MR}(\Delta \mathbf{x}_1 + \Delta \mathbf{x}_2) = \textbf{MR}(\Delta \mathbf{x}_1) + \textbf{MR}(\Delta \mathbf{x}_2).
    \]
\end{theorem}
\begin{proof}

Let $\Delta \mathbf{x}_1$ represent the implemented assignment and $\Delta \mathbf{x}_2$ represent the proposed assignment. Let $j_{\Delta \mathbf{x}_1}^+$ and $j_{\Delta \mathbf{x}_1}^-$ denote the target indices of the \textit{Designated Target ID} and \textit{Previous Target ID} for the implemented assignment, respectively. Similarly, let $j_{\Delta \mathbf{x}_2}^+$ and $j_{\Delta \mathbf{x}_2}^-$ denote the \textit{Designated Target ID} and \textit{Previous Target ID} for the proposed assignment. For instance, if the proposed assignment reallocates a weapon from target $5$ to target $7$, then $j_{\Delta \mathbf{x}_2}^+ = 7$ and $j_{\Delta \mathbf{x}_2}^- = 5$.

Recall that for any weapon-target allocation \(\mathbf{x}\), the objective function is expressed as \(J(\mathbf{x}) = T(\mathbf{x}) + C(\mathbf{x})\). Both the target survival cost \(T(\mathbf{x})\) and the weapon utilization cost \(C(\mathbf{x})\) can be decomposed into sums of terms, where each term depends solely on the allocation to a single target:
\begin{alignat*}{1}
    T(\mathbf{x}) &= \sum_{j \in \mathcal{T}} w_j P_j^\text{survive} =  \sum_{j \in \mathcal{T}} w_j\prod_{i \in \mathcal{A}} \prod_{k \in \mathcal{W}_i} (1 - p_{ikj})^{x_{ikj}}, \\
    C(\mathbf{x}) &= \sum_{i \in \mathcal{A}} \sum_{k \in \mathcal{W}_i} c_{ik} \sum_{j \in \mathcal{T}} x_{ikj} = \sum_{j \in \mathcal{T}} \sum_{i \in \mathcal{A}} \sum_{k \in \mathcal{W}_i} c_{ik} x_{ikj}.
\end{alignat*}
Thus, the total objective function \(J(\mathbf{x})\) can be expressed as a sum of functions, \(J(\mathbf{x}) = \sum_{j \in \mathcal{T}} \phi_j(\mathbf{x})\), where each \(\phi_j(\mathbf{x})\) depends only on the weapon allocation to target \(j\). 

Furthermore, observe that the modified assignment \(\mathbf{x} + \Delta \mathbf{x}_1\) differs from the previous assignment \(\mathbf{x}\) by only a single weapon reallocation (i.e., the allocation of a single weapon is changed). Let \(j_{\Delta \mathbf{x}_1}^+\) and \(j_{\Delta \mathbf{x}_1}^-\) denote the target indices of the \textit{Designated Target ID} and \textit{Previous Target ID} for the weapon being reallocated, respectively. For instance, if the third weapon of the second agent was initially assigned to target \(5\) and is reassigned to target \(7\), then \(j_{\Delta \mathbf{x}_1}^+ = 7\) and \(j_{\Delta \mathbf{x}_1}^- = 5\). 

In cases where the weapon was previously unassigned, \(j_{\Delta \mathbf{x}_1}^-\) is undefined (or empty). Conversely, if a weapon is reallocated to remain unfired, \(j_{\Delta \mathbf{x}_1}^+\) is undefined (or empty).

Thus, the total objective function $J(\mathbf{x})$ can be expressed as a sum of functions, $J(\mathbf{x}) = \sum_{j \in \mathcal{T}} \phi_j(\mathbf{x})$, where each $\phi_j(\mathbf{x})$ depends only on the weapon allocation to
target $j$. Explicitly, this function is given by:
\begin{equation}
\phi_j(\mathbf{x}) = w_j \prod_{i \in \mathcal{A}} \prod_{k \in \mathcal{W}_i}(1 - p_{ikj})^{x_{ikj}} + \sum_{i \in \mathcal{A}} \sum_{k \in \mathcal{W}_i} c_{ik} x_{ikj}.
\end{equation}
By isolating the objective into these target-specific components, we can transparently evaluate the impact of single-weapon deviations. Specifically, we may relate the initial objective function $J(\mathbf{x})$ to the modified assignment $J(\mathbf{x}+\Delta \mathbf{x}_1)$ by
\begin{alignat}{1}
    J(\mathbf{x}) = J(\mathbf{x}+\Delta \mathbf{x}_1) &+ \phi_{j_{\Delta \mathbf{x}_1}^+}(\mathbf{x}) - \phi_{j_{\Delta \mathbf{x}_1}^+}(\mathbf{x}+\Delta \mathbf{x}_1) \nonumber \\
    &+ \phi_{j_{\Delta \mathbf{x}_1}^-}(\mathbf{x}) - \phi_{j_{\Delta \mathbf{x}_1}^-}(\mathbf{x}+\Delta \mathbf{x}_1). \label{result1}
\end{alignat}
This holds because the modified allocation $\mathbf{x}+\Delta \mathbf{x}_1$ differs from the original allocation $\mathbf{x}$ only in the assignments to targets $j_{\Delta \mathbf{x}_1}^+$ and $j_{\Delta \mathbf{x}_1}^-$. Similarly, the relationship between the initial objective function $J(\mathbf{x})$ and the modified objective function $J(\mathbf{x}+\Delta \mathbf{x}_2)$, and therefore the marginal return $\textbf{MR}(\Delta \mathbf{x}_2) \coloneqq J(\mathbf{x}) - J(\mathbf{x}+\Delta \mathbf{x}_2)$, is given by
\begin{alignat}{1}
    \textbf{MR}(\Delta \mathbf{x}_2) \ = \ &\phi_{j_{\Delta \mathbf{x}_2}^+}(\mathbf{x}) - \phi_{j_{\Delta \mathbf{x}_2}^+}(\mathbf{x}+\Delta \mathbf{x}_2) \nonumber \\
    &+ \phi_{j_{\Delta \mathbf{x}_2}^-}(\mathbf{x}) - \phi_{j_{\Delta \mathbf{x}_2}^-}(\mathbf{x}+\Delta \mathbf{x}_2). \label{result2}
\end{alignat}
If \(\Delta \mathbf{x}_1\) and \(\Delta \mathbf{x}_2\) are \textbf{non-conflicting}, then none of the four target indices \(j_{\Delta \mathbf{x}_1}^+\), \(j_{\Delta \mathbf{x}_1}^-\), \(j_{\Delta \mathbf{x}_2}^+\), and \(j_{\Delta \mathbf{x}_2}^-\) overlap. Specifically, this means that: 
\begin{equation*} \scalebox{0.95}{
$j_{\Delta \mathbf{x}_1}^+ \neq j_{\Delta \mathbf{x}_2}^+, \ 
j_{\Delta \mathbf{x}_1}^+ \neq j_{\Delta \mathbf{x}_2}^-, \  
j_{\Delta \mathbf{x}_1}^- \neq j_{\Delta \mathbf{x}_2}^+, \   \text{and} \
j_{\Delta \mathbf{x}_1}^- \neq j_{\Delta \mathbf{x}_2}^-.$}\footnote{Note that in the case a target index is undefined (or empty), we do not consider two empty fields to overlap, so for our purposes define the empty set as not equal to itself.}
\end{equation*} 
Because these indices are distinct, and the total objective function can be expressed as \(J(\mathbf{x}) = \sum_{j \in \mathcal{T}} \phi_j(\mathbf{x})\), where each \(\phi_j(\mathbf{x})\) depends only on the weapon allocation to target \(j\), the combined effect of implementing both deviations \(\Delta \mathbf{x}_1\) and \(\Delta \mathbf{x}_2\) can be represented as the sum of their individual effects. Specifically:
\begin{equation}\label{result3}
\begin{split} 
    J(\mathbf{x}) = \ & J(\mathbf{x}+\Delta \mathbf{x}_1+\Delta \mathbf{x}_2)   \\
    & +  \phi_{j_{\Delta \mathbf{x}_1}^+}(\mathbf{x}) - \phi_{j_{\Delta \mathbf{x}_1}^+}(\mathbf{x}+\Delta \mathbf{x}_1) \\
    & + \phi_{j_{\Delta \mathbf{x}_1}^-}(\mathbf{x}) - \phi_{j_{\Delta \mathbf{x}_1}^-}(\mathbf{x}+\Delta \mathbf{x}_1)  \\
    & +  \phi_{j_{\Delta \mathbf{x}_2}^+}(\mathbf{x}) - \phi_{j_{\Delta \mathbf{x}_2}^+}(\mathbf{x}+\Delta \mathbf{x}_2) \\
    &+ \phi_{j_{\Delta \mathbf{x}_2}^-}(\mathbf{x}) - \phi_{j_{\Delta \mathbf{x}_2}^-}(\mathbf{x}+\Delta \mathbf{x}_2). 
\end{split}
\end{equation}
Combining the results \eqref{result1}-\eqref{result3}, this gives us $\textbf{MR}(\Delta \mathbf{x}_1 + \Delta \mathbf{x}_2) = \textbf{MR}(\Delta \mathbf{x}_1) + \textbf{MR}(\Delta \mathbf{x}_2)$ as desired. This highlights that the total impact is additive when the deviations \(\Delta \mathbf{x}_1\) and \(\Delta \mathbf{x}_2\) are \textbf{non-conflicting}, as their effects are independent and do not interfere with each other.
\end{proof}
The proof of Theorem \ref{Non-conflicting_Theorem} can be straightforwardly extended to the more general case of a set of non-conflicting assignments. In this broader context, the combined effect of implementing all assignments in the set is equal to the sum of the individual marginal returns of each assignment.
\begin{theorem}\label{Non-conflicting_Theorem_Generalization}
    Let \(\mathbf{x}\) represent the current global consensus on weapon-target allocations, and let \(\{\Delta \mathbf{x}_1, \dots , \Delta \mathbf{x}_N\}\) be a set of proposed single-weapon deviations. If every pair of deviations within this set are \textbf{non-conflicting}, then 
    \begin{equation}
        \textbf{MR}\Bigg(\sum_{r=1}^{N}\Delta \mathbf{x}_r\Bigg) = \sum_{r=1}^{N} \textbf{MR}(\Delta \mathbf{x}_r)
    \end{equation}
    where 
    \begin{alignat*}{1}
        \textbf{MR}(\Delta \mathbf{x}_r) &\coloneqq J(\mathbf{x}) - J(\mathbf{x}+\Delta \mathbf{x}_r) \\
        \textbf{MR}\Bigg(\sum_{r=1}^{N}\Delta \mathbf{x}_r\Bigg) &\coloneqq J(\mathbf{x}) - J(\mathbf{x}+\sum_{r=1}^{N}\Delta \mathbf{x}_r) \\
    \end{alignat*}
    denote the individual and combined effects of modifying the current assignment \(\mathbf{x}\) by \(\Delta \mathbf{x}_r\) and the entire set of deviations, respectively.
\end{theorem}
\begin{proof}
The proof generalizes the reasoning in Theorem \ref{Non-conflicting_Theorem} by induction on the number of deviations \(N\). First, recall that the objective function \(J(\mathbf{x}) = \sum_{j \in \mathcal{T}} \phi_j(\mathbf{x})\) decomposes into target-specific terms, where each \(\phi_j(\mathbf{x})\) depends only on the weapons allocated to target \(j\). Each single-weapon deviation \(\Delta \mathbf{x}_r\) modifies at most two targets: the previous target \(j_{\Delta \mathbf{x}_r}^-\) (if the weapon was allocated) and the new target \(j_{\Delta \mathbf{x}_r}^+\). By the non-conflicting condition, all \(j_{\Delta \mathbf{x}_r}^+\) and \(j_{\Delta \mathbf{x}_r}^-\) across \(r = 1, \dots, N\) are distinct. Thus, applying all deviations collectively alters each affected target exactly once. Consequently, the total marginal return is the sum of individual returns:
\begin{equation*}
    \begin{split}
        &\textbf{MR}\left(\sum_{r=1}^{N}\Delta \mathbf{x}_r\right) \\
        & \scalebox{0.80}{$= \sum_{r=1}^{N} \left[\phi_{j_{\Delta \mathbf{x}_r}^+}(\mathbf{x}) - \phi_{j_{\Delta \mathbf{x}_r}^+}(\mathbf{x}+\Delta \mathbf{x}_r) + \phi_{j_{\Delta \mathbf{x}_r}^-}(\mathbf{x}) - \phi_{j_{\Delta \mathbf{x}_r}^-}(\mathbf{x}+\Delta \mathbf{x}_r)\right] $}\\
        &= \sum_{r=1}^{N} \textbf{MR}(\Delta \mathbf{x}_r).
    \end{split}
\end{equation*}
\end{proof}
\begin{corollary}\label{cor:nonconflicting_parallel}
As a consequence of Theorem \ref{Non-conflicting_Theorem_Generalization}, we may asses the impact of a single-weapon deviation from the current global consensus independently of other deviations, with the assurance that since only \textbf{non-conflicting} bids are accepted in the Consensus phase of DMR, the additive property of marginal returns guarantees that the net improvement to the objective function from simultaneously implementing all accepted bids equals the arithmetic sum of their individual marginal returns. This independence eliminates the need for sequential bid validation and ensures the integrity of locally computed marginal returns is preserved, even when multiple agents propose concurrent deviations.
\end{corollary}
The complete pseudocode for the DMR algorithm is provided in Algorithms \ref{alg:dmr}, \ref{alg:queue}, and \ref{alg:remove}. This implementation assumes that every agent in the battlespace has a direct communication link with every other agent (i.e., the communication network is strongly connected) and that the algorithm runs for a predefined number of inter-agent communication rounds, $P$. The termination condition in Algorithm \ref{alg:dmr} can be trivially modified to instead halt only when the agents have converged to a solution (when the aggregate queue in line 10 of Algorithm \ref{alg:dmr} is empty), or to terminate as soon as either we reach the $P+1$ phase or the solution has converged, whichever comes first.

Notably, the DMR implementation adheres to the design constraints established earlier. Crucially, agent $i$ computes its marginal returns by evaluating its own weapon capabilities against its locally maintained consensus state $\mathbf{x}_i(t)$, circumventing the need for global sensory observation of other agents' real-time states. In Algorithm \ref{alg:queue}, each agent incorporates only the pertinent weapon-target characteristics into its bid, ensuring that decisions are based solely on local weapon attributes. The only information shared across the battlespace is limited to target-specific data (unique target IDs and associated values), thereby maintaining the confidentiality of each agent’s capabilities. This careful partitioning of local and global information guarantees that the algorithm fulfills the intended decentralized communication and decision-making requirements.

\begin{algorithm}[htbp]
\small
\caption{Distributed Marginal Return (DMR) Algorithm}\label{alg:dmr}
\begin{algorithmic}[1]
\Require Agents $\mathcal{A}$, targets $\mathcal{T}$, weapon kill probabilities $p_{ikj}$, weapon utilization costs $c_{ik}$, target weights $w_j$, maximum number of consensus phases $P$
\State \textbf{Initialization:} For each agent $i\in \mathcal{A}$, set current assignment $\mathbf{x}_i \gets \mathbf{0}$.
\For{$phase = 1,2,\dots,P$}

    \noindent \Comment{Auction Phase \& Network Relay}
    \For{each agent $i\in \mathcal{A}$}
        \State $Q_i \gets \Call{GenerateQueue}{\mathbf{x}_i,i}$ 
        \For{each spatial neighbour $j\in \mathcal{N}_i$}
            \State Share $Q_i$ and relay any previously received queues to neighbour $j$
        \EndFor
    \EndFor
    
    \Comment{Consensus Phase}
    \For{each agent $i\in \mathcal{A}$}
        \State \textbf{Wait} until queues from all agents in $\mathcal{A}$ are successfully received
        \State $Q \gets \bigcup_{j\in \mathcal{A}} Q_j$ \Comment{Aggregate all proposed bids}
        \State Sort $Q$ in descending order by marginal return.
        \While{$Q \neq \emptyset$}
            \State \textbf{Remove} the first bid $a$ from $Q$
            \State \textbf{Implement} bid $a$ (ie update current assignment $\mathbf{x}_i$ according to bid $a$)
            \State $Q \gets \Call{RemoveConflictingBids}{Q,a}$ 
        \EndWhile
    \EndFor
\EndFor
\State \Return Final weapon-target assignments $\mathbf{x}_i$
\end{algorithmic}
\end{algorithm}

\begin{algorithm}[htbp]
\small
\caption{GenerateQueue for Agent $i$}\label{alg:queue}
\begin{algorithmic}[1]
\Require Current assignment $\mathbf{x}_i$, weapon set $\mathcal{W}_i$, targets $\mathcal{T}$, kill probabilities $p_{ikj}$, weapon costs $c_{ik}$, target weights $w_j$
\State Compute the current objective cost $J(\mathbf{x}_i)$
\State Initialize an empty queue $Q$
\For{each weapon $k\in \mathcal{W}_i$}
    \State Let $j_{\text{curr}}$ be the current target (if any) to which weapon $k$ is assigned.
    \For{each target $j\in \mathcal{T} \cup \{\emptyset\}$ such that $j\neq j_{\text{curr}}$}
        \State Form a \emph{proposed assignment} that reallocates weapon $k$ to target $j$
        \State Compute the corresponding marginal return $\Delta = J(\mathbf{x}_i) - J(\mathbf{x}_i + \Delta \mathbf{x})$
        \If{$\Delta > 0$}
            \State Append \emph{proposed assignment} to \(Q\), where
            \[
            \emph{proposed assignment} = \left\{
            \begin{array}{ll}
                \text{agent\_id}: & i,\\[0.2em]
                \text{weapon\_id}: & k,\\[0.2em]
                \text{designated\_target\_id}: & j,\\[0.2em]
                \text{previous\_target\_id}: & j_{\text{curr}},\\[0.2em]
                \text{probability\_of\_kill}: & p_{ikj},\\[0.2em]
                \text{weapon\_utilization\_cost}: & c_{ik},\\[0.2em]
                \text{marginal\_return}: & \Delta
            \end{array}
            \right\}
            \]
        \EndIf
    \EndFor
\EndFor
\State Sort $Q$ in descending order by $\Delta$
\State \Return $Q$
\end{algorithmic}
\end{algorithm}

\begin{algorithm}[htbp]
\small
\caption{RemoveConflictingBids}\label{alg:remove}
\begin{algorithmic}[1]
\Require Aggregate queue \( Q \) of bids, implemented bid \( a \) with fields:
\begin{itemize}
    \item \( a.\mathtt{agent\_id} \), \( a.\mathtt{weapon\_id} \)
    \item \( a.\mathtt{designated\_target\_id} \), \( a.\mathtt{previous\_target\_id} \)
\end{itemize}
\State Initialize new queue \( Q' \gets \emptyset \)
\For{each bid \( b \in Q \) with fields:
\begin{itemize}
    \item \( b.\mathtt{agent\_id} \), \( b.\mathtt{weapon\_id} \)
    \item \( b.\mathtt{designated\_target\_id} \), \( b.\mathtt{previous\_target\_id} \)
\end{itemize}}
    \If{\textbf{NOT} \big[\((b.\mathtt{agent\_id} = a.\mathtt{agent\_id} \) \textbf{and} \( b.\mathtt{weapon\_id} = a.\mathtt{weapon\_id})\)\big] \textbf{and} 
    \textbf{NOT} \bigg[\(\Call{Match}{b.\mathtt{designated\_target\_id}, a.\mathtt{designated\_target\_id}}\) \newline \textbf{or} \ \ \(\Call{Match}{b.\mathtt{designated\_target\_id}, a.\mathtt{previous\_target\_id}}\) \newline \textbf{or} \ \ \(\Call{Match}{b.\mathtt{previous\_target\_id}, a.\mathtt{designated\_target\_id}}\) \newline \textbf{or} \ \ \(\Call{Match}{b.\mathtt{previous\_target\_id}, a.\mathtt{previous\_target\_id}}\)\bigg]}
        \State Append \( b \) to \( Q' \)
    \EndIf
\EndFor
\State \Return \( Q' \)
\newline
\noindent\textbf{Helper Function:} \textsc{Match}(\( x, y \)) returns \textbf{True} if \( x = y \), \textbf{False} otherwise; with the exception that \textsc{Match}(\( \emptyset, \emptyset \)) returns \textbf{False}. 
\end{algorithmic}
\end{algorithm}

\subsection{Computational Complexity Analysis}
In this section, we analyze the computational complexity of the Distributed Marginal Return (DMR) algorithm in terms of three key parameters: the number of agents, the number of weapons per agent, and the number of targets. Let \(A\) denote the number of agents in the battlespace, \(W\) the number of weapons available per agent (noting that the DMR algorithm is capable of handling heterogeneous weapon counts, but for analytical convenience we assume each agent has exactly \(W\) weapons), and \(T\) the total number of targets.

The algorithm comprises two principal phases: the Auction phase and the Consensus phase. During the Auction phase (see Algorithm~\ref{alg:queue}), each agent independently constructs a queue of potential weapon-target reassignments. For every agent, the algorithm iterates over each of its \(W\) weapons and, for each weapon, considers reassigning it to each of the \(T\) targets not currently engaged. The evaluation of each potential reassignment involves computing the marginal return by updating a global objective function that aggregates contributions from all agents. This evaluation has a cost proportional to the number of agents and weapons, yielding a per-agent complexity of \(O(W \times T \times AW) = O(AW^2T)\). 

The Consensus phase (detailed in Algorithms~\ref{alg:dmr} and \ref{alg:remove}) further affects the computational cost. In this phase, each agent aggregates the bid queues received from all agents into a single global queue that, in the worst-case scenario, may contain \(O(AWT)\) bids. Sorting this aggregate queue requires \(O(AWT \log (AWT))\) time. Moreover, as each bid is implemented, the algorithm must compare it with all remaining bids to remove those that conflict—a process that can require up to \(O(W \times AWT)\) comparisons per agent. This results in a per-agent complexity of \(O(AWT(W+\log (AWT)))\) for the Consensus phase.

\begin{table}[h]
\centering
\caption{Computational complexity of the DMR algorithm. \(A\): agents, \(W\): weapons/agent, \(T\): targets.}
\label{tab:complexity}
\begin{tabular}{@{}ll@{}}
\toprule
\textbf{Aspect} & \textbf{Complexity} \\
\midrule
Auction phase (time) & \(O(AW^2T)\) \\
Consensus phase (time) & \(O(AWT(W + \log(AWT)))\) \\
Total time complexity & \(O(AW^2T + AWT\log(AWT))\) \\
Space complexity & \(O(AWT)\) \\
\bottomrule
\end{tabular}
\end{table}

Overall, the per-phase time complexity of the DMR algorithm is dominated by the combined cost of the Auction and Consensus phases, expressed as \(O(AW^2T + AWT\log (AWT))\). This indicates that the algorithm scales quadratically with respect to the number of weapons per agent \(W\), while exhibiting log-linear scaling with respect to the number of agents \(A\) and targets \(T\).

In terms of space complexity, each agent maintains data structures for current assignments and proposed bids. The space required for current assignments is $O(AW)$ per agent, while for proposed bids, it is $O(AWT)$. Thus, the overall space complexity is $O(AWT)$.

This complexity analysis demonstrates that while the DMR algorithm exhibits quadratic scaling in the number of weapons, its design effectively leverages distributed computation, making it a practical approach for decentralized weapon-target assignment.

\begin{table}[t]
\centering
\caption{Per-agent time and space complexity of distributed algorithms, per inter-agent communication round.}
\resizebox{1.0\columnwidth}{!}{
\begin{tabular}{lcc}
\toprule
\textbf{Algorithm} & \textbf{Time Complexity} & \textbf{Space Complexity} \\
\midrule
DMR   & $O(AW^2T + AWT\log(AWT))$ & $O(AWT)$ \\
CBAA  & $O(AWT)$                 & $O(WT)$  \\
GRAPE & $O(W(A + T))$            & $O(AWT)$ \\
\bottomrule
\end{tabular}
}
\label{tab:complexity_summary}
\end{table}

\section{Numerical Experiments and Results}
To evaluate the effectiveness of our proposed Distributed Marginal Return (DMR) algorithm, we conducted a series of Monte Carlo simulation studies to benchmark its performance against both distributed algorithms and globally optimal solutions set in low problem dimensions. 

The Monte Carlo simulation studies were designed to address three key questions. \textit{First}, which of the distributed algorithms achieves the best weapon-target allocation for a given number of inter-agent communication rounds? \textit{Second}, how closely do these algorithms approach the globally optimal solution in small-scale scenarios? \textit{Third}, what is the scalability of these algorithms with respect to increasing problem complexity? %, again normalizing for inter-agent communication rounds. 
%with respect to increasing (i) the number of targets, (ii) the number of agents, and (iii) the number of weapons per agent. 
To address the questions, we conducted the evaluation studies summarised below:

\begin{itemize}
    \item The first study compared DMR to two established distributed algorithms: the Consensus-based Auction Algorithm (CBAA) \cite{CBAA} and the GRAPE algorithm \cite{GRAPE} (see Section~\ref{sec:comp-methods}).
    \item The second, smaller-scale study introduced a baseline computed via Full Enumeration (FE), enabling comparison with the provably optimal allocation. This normalization is necessary because all three algorithms are computationally efficient and do not exhibit exponential complexity.
    \item The third study evaluated the scalability of all three distributed algorithms by systematically varying key problem parameters: (i) the number of targets, (ii) the number of agents, and (iii) the number of weapons per agent under settings of limited communication rounds. By systematically varying these parameters, we assess how well each algorithm adapts to larger and more complex problem instances while maintaining efficient communication overhead. 
\end{itemize}
%These studies provide a comprehensive assessment of how each algorithm adapts to increasing problem complexity while maintaining efficient communication overhead. %Specifically, we analyze how performance is affected when increasing (i) the number of targets, (ii) the number of agents, and (iii) the number of weapons per agent. 
To rigorously evaluate and compare the algorithms, we establish two primary representative indicators for performance. First, \textit{solution quality} is quantified using the total assignment cost (the objective function $J(\mathbf{x})$). This provides a unified metric that captures both the expected survival value of the targets and the utilization cost of the deployed weapons, with lower values indicating a superior and more resource-efficient allocation. Second, \textit{communication efficiency} is measured by the number of inter-agent communication rounds required to reach a solution. In decentralized, contested environments, the computational time at individual agents is typically negligible compared to the latency of wireless transmission. Therefore, minimizing communication rounds is the most critical metric for reducing system latency, conserving limited bandwidth, and maintaining operational security by limiting electromagnetic emissions.

\subsection{Settings and Comparison Methods}\label{sec:comp-methods}
The primary bottleneck in distributed execution is the latency associated with inter-agent communication, specifically, the time spent broadcasting and receiving messages before proceeding to the next phase of the algorithm. Furthermore, as discussed in Section \ref{section_dmr}, minimizing inter-agent communication is crucial for operational security, as electromagnetic emissions can reveal agent positions, increasing the risk of detection and localization by adversaries. Given these considerations, we evaluate performance in relation to the number of inter-agent communication rounds.

While CBAA was originally designed for the agent-task allocation problem rather than the weapon-target assignment formulation described in Section \ref{section_problem_formulation}, we developed a conversion process to facilitate a meaningful comparison. In this adaptation, each weapon was treated as an independent agent, and each target was mapped to a task. Additional modifications were made to accommodate the fact that our objective function in \eqref{min_J_x} is a minimization rather than a maximization problem. However, a key limitation of this conversion is that CBAA enforces a \textit{conflict-free} assignment, meaning that each task (or target, in our case) can be assigned to at most one agent (weapon). It is important to note that CBAA's structural limitation to one-to-one assignments places it at an inherent disadvantage for WTA; illustrating the impact of this exact limitation serves as the primary motivation for the development of DMR's multi-weapon conflict-resolution protocol.

%In contrast, the weapon-target assignment problem allows multiple weapons to engage a single target. Due to the fundamental structure of CBAA’s auction and consensus phases, it was not possible to relax this conflict-free constraint without significantly altering the algorithm’s mechanics, which limits its applicability to our problem domain.

In contrast, the GRAPE algorithm, a distributed multi-agent approach inspired by game theory, was designed for multi-robot task allocation. Unlike CBAA, GRAPE could be faithfully adapted to the weapon-target assignment problem without requiring modifications that introduce artificial constraints or deviations from the original problem formulation.

\begin{figure}[h]
\centering
\includegraphics[trim=20 20 0 0, clip, width=0.8\columnwidth]{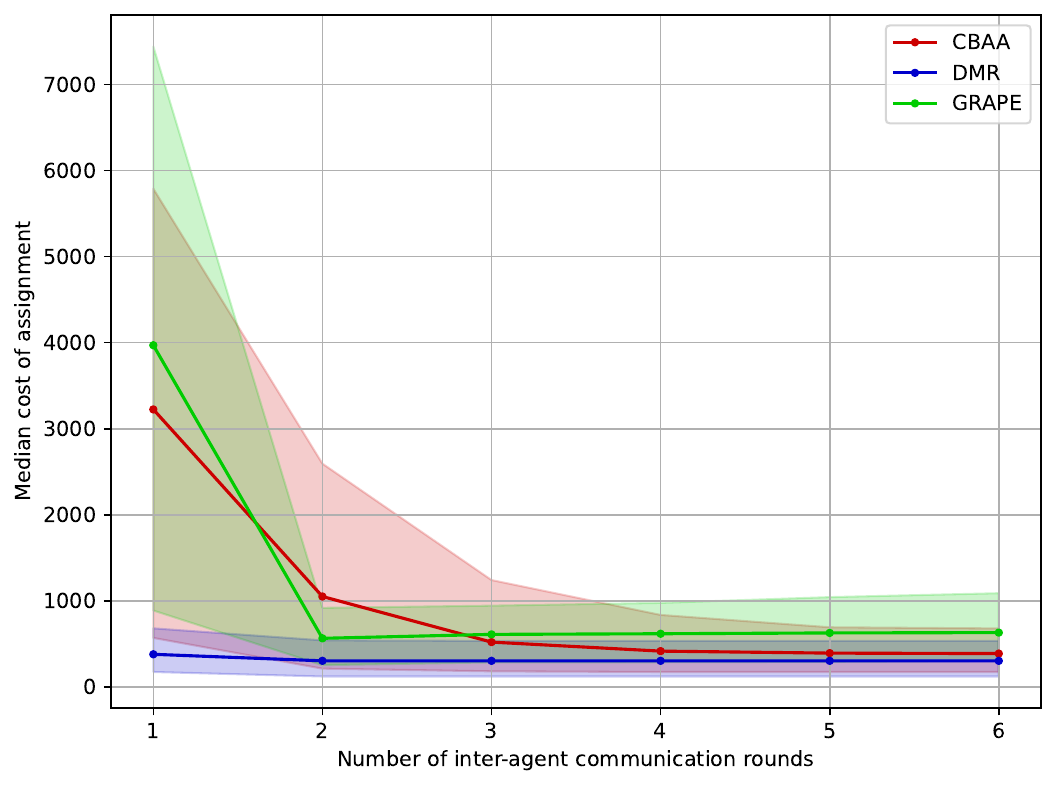}
\caption{Comparison of the median objective function value across inter-agent communication rounds for the Distributed Marginal Return (DMR), Consensus-based Auction Algorithm (CBAA), and GRAPE algorithms. Shaded regions represent the 25th to 75th interquartile range (IQR) across 100 Monte Carlo simulations.}
\label{fig:sim1}
\end{figure}

\subsection{Overall Performance}
In the first Monte Carlo simulation study (Figure \ref{fig:sim1}), we evaluate the performance of our proposed Distributed Marginal Return (DMR) algorithm relative to the Consensus-based Auction Algorithm (CBAA) and the GRAPE algorithm. For each of the 100 simulation runs, a random weapon-target assignment (WTA) problem is generated by uniformly sampling the number of targets (ranging from 1 to 30), the number of agents (ranging from 1 to 15), and the number of weapons per agent (ranging from 1 to 10). In addition, all associated parameters, including target priority weights and weapon-target kill probabilities, are randomly generated for each instance. Each simulation instance is then executed for a fixed number of inter-agent communication rounds (up to 6 rounds), and the objective function value, as defined in Equation \eqref{J_x}, is recorded after each round.

The performance of each algorithm is assessed by aggregating the results over all simulation instances. For every communication round, we compute the median objective cost and the interquartile range (IQR) across all simulations, providing robust performance metrics that account for the inherent variability in the problem parameters. The resulting performance curves, plotted with the number of inter-agent communication rounds on the x-axis and the median assignment cost on the y-axis, with shaded regions indicating the IQR, facilitate a direct comparison of convergence speed and solution quality among the three distributed algorithms. The results in Figure \ref{fig:sim1} clearly show that the DMR algorithm achieves a significantly lower median assignment cost within the first communication round than its counterparts. While the performance of all three algorithms improves with additional communication rounds, DMR consistently maintains its superiority, converging to a better-quality solution much more rapidly. This normalization to communication rounds is particularly pertinent given that, in distributed environments, the primary bottleneck is the latency incurred during inter-agent communications rather than the computational cost at each agent.

\begin{figure}[h]
\centering
\includegraphics[trim=20 20 0 0, clip, width=0.8\columnwidth]{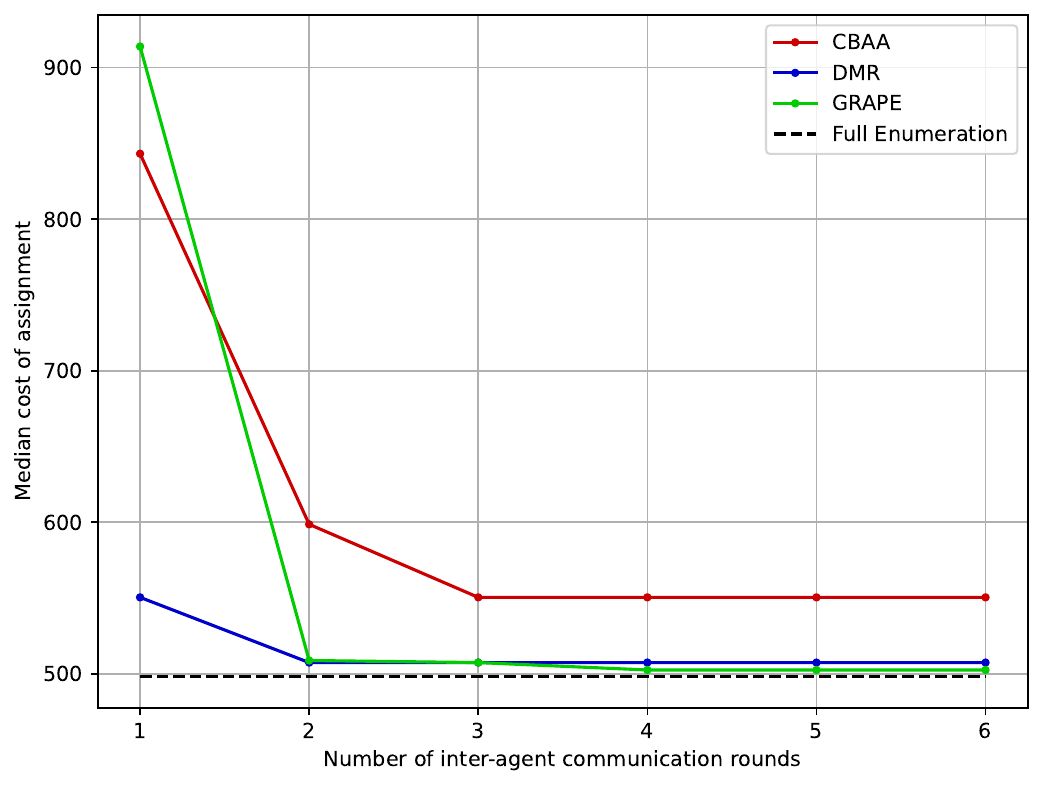}
\caption{Comparison of the median objective function value across inter-agent communication rounds for the Distributed Marginal Return (DMR), Consensus-based Auction Algorithm (CBAA), and GRAPE algorithms. Black-dashed line denotes the provably optimal solution computed using Full Enumeration.}
\label{fig:sim_small1}
\end{figure}

\subsection{Comparison Case Study with Globally Optimal Solutions}
In the second Monte Carlo study (Figure \ref{fig:sim_small1}), we analyzed small-scale problems to compare the distributed algorithms against the globally optimal solution obtained via Full Enumeration (FE). Due to FE’s exponential computational complexity, problem parameters were restricted to: 2–5 targets, 1–3 agents, and 1–3 weapons per agent. For each of 100 simulations, we generated random WTA instances within these bounds, executed the algorithms for up to 6 communication rounds, and recorded the median assignment cost. Unlike the first study, we omitted IQR visualization and instead plotted FE’s median assignment cost as a horizontal black dashed line. Since FE computes the global minimum in a centralized manner without requiring inter-agent communication, its performance remains constant across all rounds, serving as an optimality benchmark. This comparison reveals how closely the distributed algorithms approach the theoretical optimum under constrained problem scales where FE is computationally tractable. The results show that DMR converges to a solution nearly identical to the global optimum within just one to two communication rounds. In contrast, CBAA and GRAPE require more rounds and converge to a less optimal solution, highlighting DMR's efficiency and effectiveness in finding high-quality assignments. 

\begin{figure}[h]
\centering
\includegraphics[trim=20 20 0 0, clip, width=0.8\columnwidth]{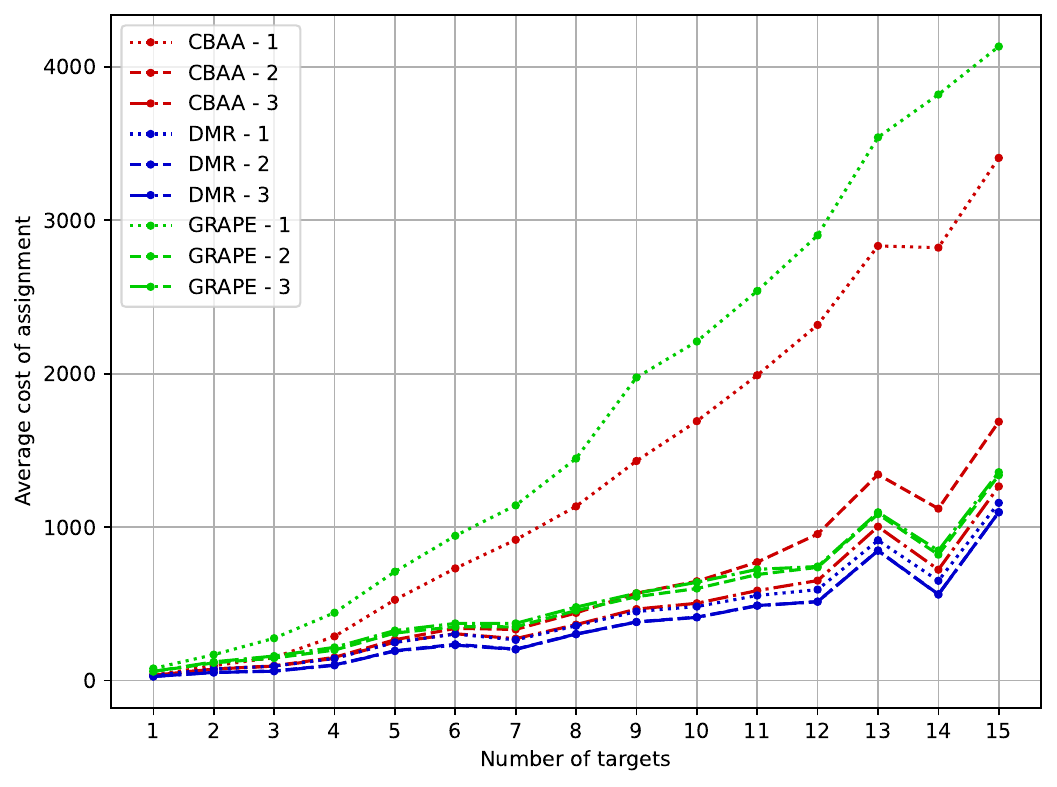}
\caption{Median assignment cost over 1 to 3 communication rounds, specified by the legend number, as a function of the number of targets.}
\label{fig:sim2}
\end{figure}

\begin{figure}[h]
\centering
\includegraphics[trim=20 20 0 0, clip, width=0.8\columnwidth]{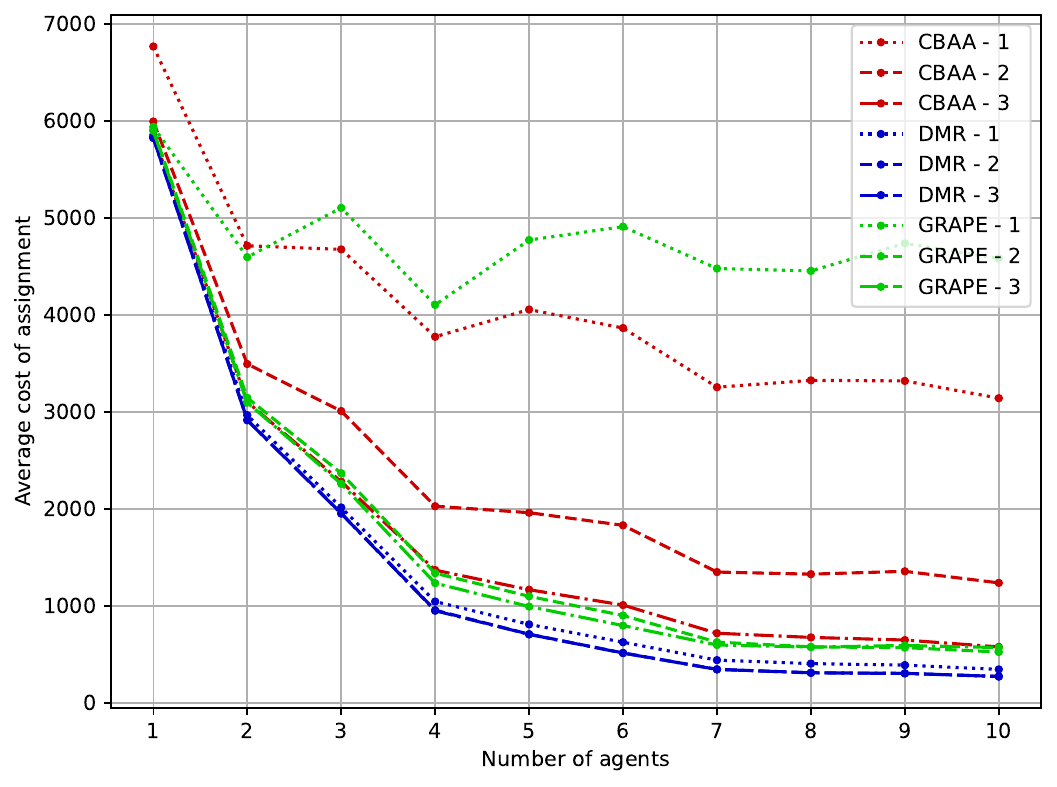}
\caption{Median assignment cost over 1 to 3 communication rounds, specified by the legend number, as a function of the number of agents.}
\label{fig:sim3}
\end{figure}

\begin{figure}[h]
\centering
\includegraphics[trim=20 20 0 0, clip, width=0.8\columnwidth]{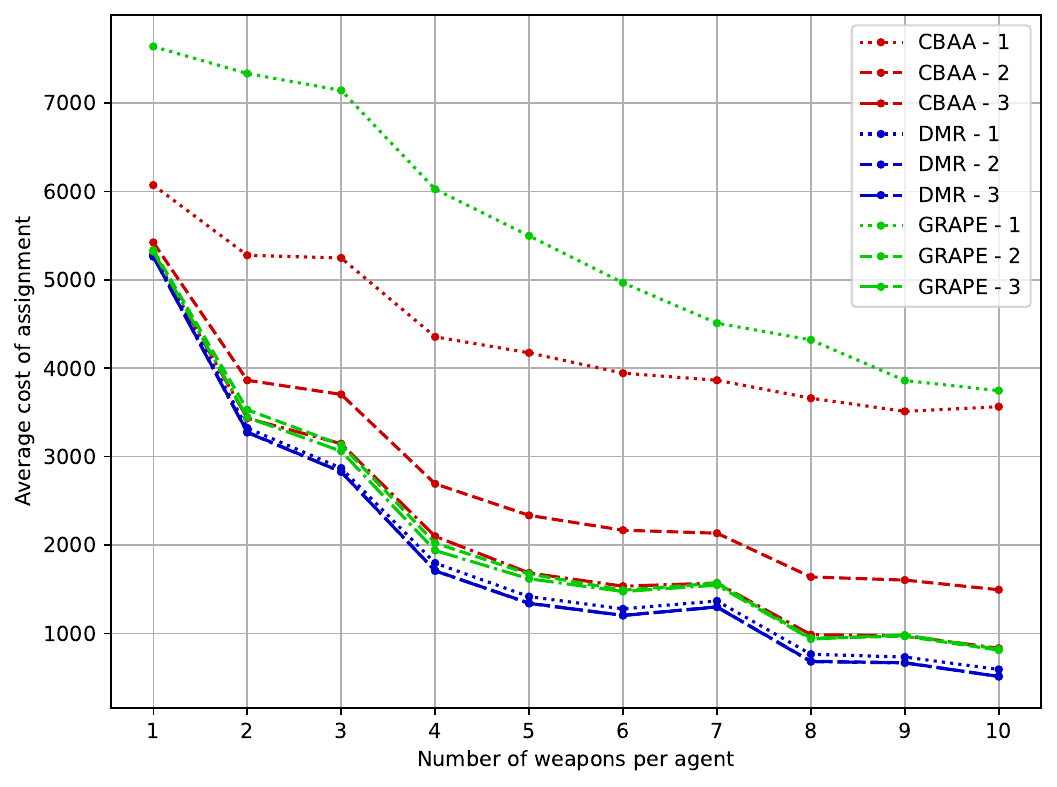}
\caption{Median assignment cost over 1 to 3 communication rounds, specified by the legend number, as a function of the number of weapons per agent.}
\label{fig:sim4}
\end{figure}

\subsection{Scalability Under Limited Communication Rounds}
The next three Monte Carlo simulation studies evaluate the scalability of the Distributed Marginal Return (DMR), Consensus-Based Auction Algorithm (CBAA), and GRAPE algorithms. Specifically, we analyze how assignment cost is affected when systematically varying (i) the number of targets, (ii) the number of agents, and (iii) the number of weapons per agent. As in previous studies, we recorded the mean assignment cost over 100 Monte Carlo trials, evaluating performance after one, two, and three inter-agent communication rounds. The average cost for each algorithm at each round is visualized as distinct but consistently coloured lines, enabling direct comparison of convergence behaviour and solution quality under increasing problem complexity.  

In the first of these three studies (Figure \ref{fig:sim2}), we examine how performance scales with the number of targets. Each problem instance is generated by uniformly sampling the number of agents (ranging from 1 to 15) and the number of weapons per agent (ranging from 1 to 10), while the number of targets is deterministically set between 1 and 15. For each fixed target count, 100 problem instances are generated, with target priority weights and weapon-target kill probabilities sampled randomly. The results, plotted as a function of the number of targets, provide insight into how each algorithm adapts to increasing problem dimensionality and whether additional targets disproportionately impact assignment costs across different solution approaches.  

The second and third studies follow a similar methodology but vary different parameters. In Figure \ref{fig:sim3}, we deterministically set the number of agents between 1 and 10, generating 100 Monte Carlo instances for each agent count, while the number of targets and the number of weapons per agent are randomly generated as in the previous study. The third study (Figure \ref{fig:sim4}) randomly selects between 1 and 30 targets and between 1 and 10 weapons per agent, while deterministically varying the number of weapons per agent to assess how surplus resources influence performance. In each case, the x-axis corresponds to the deterministically varied parameter, while the y-axis records the mean assignment cost at different communication rounds. These studies build upon the previous analyses, providing a comprehensive evaluation of the computational efficiency and solution quality of each algorithm under varying problem scales, with direct implications for large-scale weapon-target assignment scenarios.

The findings presented in Figures \ref{fig:sim2}, \ref{fig:sim3}, and \ref{fig:sim4} consistently demonstrate that DMR outperforms both CBAA and GRAPE across all tested scales, including varying numbers of targets, agents, and weapons per agent. In every scenario, DMR achieves a lower median assignment cost, reinforcing its superior scalability and robustness.

\subsection{Summary}
The results of all four Monte Carlo studies demonstrate the superior performance of the Distributed Marginal Return (DMR) algorithm in achieving a low-cost weapon-target assignment with significantly fewer communication rounds compared to the Consensus-Based Auction Algorithm (CBAA) and the GRoup Agent Partitioning and placing Event (GRAPE) algorithm. In most scenarios, DMR required only a single communication round to reach an assignment cost that CBAA and GRAPE could only approximate after 3–6 rounds of communication. This trend was consistent across varying numbers of targets, agents, and weapons per agent, highlighting DMR’s robustness in different problem configurations. The efficiency of DMR in rapidly converging to a high-quality solution underscores its advantage in real-time, distributed decision-making environments where minimizing communication overhead is critical. These results affirm DMR’s potential for scalable and communication-efficient deployment in dynamic scenarios, where rapid coordination and optimal resource utilization are paramount.

\section{Conclusion}
This paper introduced the Distributed Marginal Return (DMR) algorithm, a decentralized approach to the multi-agent weapon target assignment (WTA) problem. By leveraging a two-phase marginal return and conflict resolution protocol, DMR successfully overcomes the limitations of existing distributed algorithms in evaluating the coupled effects of multi-weapon assignments. Extensive Monte Carlo simulations and rigorous theoretical analyses confirm that DMR achieves rapid convergence and scales effectively with problem size. Notably, the algorithm converges to near-optimal, globally consistent assignments within 1–3 communication rounds, significantly outperforming baseline approaches like CBAA and GRAPE while minimizing inter-agent communication overhead. Future work will extend DMR to dynamic WTA scenarios and integrate sensor networks for adaptive target prioritization, further advancing resilient and scalable decision-making frameworks for modern autonomous defence systems.

%\bibliographystyle{cas-model2-names}

\section*{Acknowledgments}
This work was supported by the Defence Science and Technology Group (DSTG), Australia.

\bibliographystyle{IEEEtran}
\bibliography{sample-base}

@ARTICLE{ZhiLin2025,
  author={Lin, Zhi and Feng, Zimo and Guo, Kefeng and Nauman, Ali and Niyato, Dusit and Wang, Jiangzhou},
  journal={IEEE Wireless Communications}, 
  title={AI-Driven Seamless and Massive Access in Space-Air-Ground Integrated Networks}, 
  year={2025},
  volume={32},
  number={3},
  pages={72-79},
  doi={10.1109/MWC.001.2400371}}

@article{Zimo2026,
author = {Feng, Zimo and Lin, Zhi and Wang, Hongjun and Ma, Ruiqian and An, Kang and He, Yuanzhi},
year = {2026},
pages = {4848-4860},
title = {Toward Secure and Reliable SAGIN: Learning-Driven Multi-Dimensional Resource Scheduling for Multi-RIS-Assisted OTFS Transmission},
volume = {44},
journal = {IEEE Journal on Selected Areas in Communications},
doi = {10.1109/JSAC.2026.3700114}
}

@ARTICLE{Yuanzhi2022,
  author={He, Yuanzhi and Sheng, Biao and Yin, Hao and Yan, Di and Zhang, Yingchao},
  journal={China Communications}, 
  title={Multi-objective deep reinforcement learning based time-frequency resource allocation for multi-beam satellite communications}, 
  year={2022},
  volume={19},
  number={1},
  pages={77-91},
  doi={10.23919/JCC.2022.01.007}}

@article{Hendrickson2021,
  title={Totally Asynchronous Primal-Dual Convex Optimization in Blocks},
  author={Katherine R. Hendrickson and Matthew T. Hale},
  journal={IEEE Transactions on Control of Network Systems},
  year={2021},
  volume={10},
  pages={454-466},
  url={https://api.semanticscholar.org/CorpusID:221376866}
}

@Article{aerospace2022,
AUTHOR = {Li, Weihua and Lyu, Yongxi and Dai, Sifan and Chen, Huakun and Shi, Jingping and Li, Yongfeng},
TITLE = {A Multi-Target Consensus-Based Auction Algorithm for Distributed Target Assignment in Cooperative Beyond-Visual-Range Air Combat},
JOURNAL = {Aerospace},
VOLUME = {9},
YEAR = {2022},
NUMBER = {9},
DOI = {10.3390/aerospace9090486}
}

@ARTICLE{Xin2019,
  author={Xin, Bin and Wang, Yipeng and Chen, Jie},
  journal={IEEE Transactions on Systems, Man, and Cybernetics: Systems}, 
  title={An Efficient Marginal-Return-Based Constructive Heuristic to Solve the Sensor–Weapon–Target Assignment Problem}, 
  year={2019},
  volume={49},
  number={12},
  pages={2536-2547},
  doi={10.1109/TSMC.2017.2784187}}

@ARTICLE{CBAA,
  author={Choi, Han-Lim and Brunet, Luc and How, Jonathan P.},
  journal={IEEE Transactions on Robotics}, 
  title={Consensus-Based Decentralized Auctions for Robust Task Allocation}, 
  year={2009},
  volume={25},
  number={4},
  pages={912-926},
  doi={10.1109/TRO.2009.2022423}}

@ARTICLE{GRAPE,
  author={Jang, Inmo and Shin, Hyo-Sang and Tsourdos, Antonios},
  journal={IEEE Transactions on Robotics}, 
  title={Anonymous Hedonic Game for Task Allocation in a Large-Scale Multiple Agent System}, 
  year={2018},
  volume={34},
  number={6},
  pages={1534-1548},
  doi={10.1109/TRO.2018.2858292}}

@ARTICLE{Ahuja20071136,
author={Ahuja, R.K. and Kumar, A. and Jha, K.C. and Orlin, J.B.},
title={Exact and heuristic algorithms for the weapon-target assignment problem},
journal={Operations Research},
year={2007},
volume={55},
number={6},
pages={1136-1146},
doi={10.1287/opre.1070.0440},
}

@ARTICLE{Andersen2022,
author = {Andersen, Alexandre and Pavlikov, Konstantin and Toffolo, Túlio},
year = {2022},
month = {05},
title = {Weapon-Target Assignment Problem: Exact and Approximate Solution Algorithms},
volume = {312},
journal = {Annals of Operations Research},
doi = {10.1007/s10479-022-04525-6}
}

@ARTICLE{Boardman2017417,
author={Boardman, N.T. and Lunday, B.J. and Robbins, M.J.},
title={Heterogeneous surface-to-air missile defense battery location: a game theoretic approach},
journal={Journal of Heuristics},
year={2017},
volume={23},
number={6},
pages={417-447},
doi={10.1007/s10732-017-9350-0}
}

@ARTICLE{Chen20092006,
author={Chen, J. and Xin, B. and Peng, Z.H. and Dou, L.H. and Zhang, J.},
title={Evolutionary decision-makings for the dynamic weapon-target assignment problem},
journal={Science in China Series F: Information Sciences},
year={2009},
volume={52},
number={11},
pages={2006-2018}
}

@ARTICLE{Denbroeder1959322,
author={Denbroeder, G.G. and Ellison, R.E. and Emerling, L.},
title={On optimum target assignments},
journal={Operations Research},
year={1959},
volume={7},
number={3},
pages={322-326}
}

@CONFERENCE{Johansson2010,
author={Johansson, F. and Falkman, G.},
title={Real-time allocation of defensive resources to rockets, artillery, and mortars},
journal={13th Conference on Information Fusion, Fusion 2010},
year={2010},
doi={10.1109/icif.2010.5712026}
}

@ARTICLE{Kline20171,
author={Kline, A. and Ahner, D. and Lunday, B.},
title={Real-time heuristic algorithms for the static weapon target assignment problem},
journal={Journal of Heuristics},
year={2019},
volume={25},
pages={377–397},
doi={10.1007/s10732-018-9401-1}
}

@ARTICLE{Kwon1999640,
author={Kwon, O. and Kang, D. and Lee, K. and Park, S.},
title={Lagrangian relaxation approach to the targeting problem},
journal={Naval Research Logistics},
year={1999},
volume={46},
number={6},
pages={640-653}
}

@ARTICLE{Lee2002287,
author={Lee, Z.-J. and Su, S.-F. and Lee, C.-Y.},
title={A genetic algorithm with domain knowledge for weapon-target assignment problems},
journal={Journal of the Chinese Institute of Engineers, Transactions of the Chinese Institute of Engineers, Series A/Chung-kuo Kung Ch'eng Hsuch K'an},
year={2002},
volume={25},
number={3},
pages={287-295},
doi={10.1080/02533839.2002.9670703}
}

@article{Lee2002,
title = {An immunity-based ant colony optimization algorithm for solving weapon–target assignment problem},
journal = {Applied Soft Computing},
volume = {2},
number = {1},
pages = {39-47},
year = {2002},
doi = {https://doi.org/10.1016/S1568-4946(02)00027-3},
author = {Zne-Jung Lee and Chou-Yuan Lee and Shun-Feng Su}
}

@Article{Manne1958,
author={Manne, A.S.},
title={A target-assignment problem},
journal={Operations Research},
year={1958},
volume={6},
number={3},
pages={346-351}
}

@ARTICLE{Metler1990,
author={Metler, W.A. and Preston, F.L. and Hofmann, J.},
title={A suite of weapon assignment algorithms for a SDI mid-course battle manager},
journal={A Suite of Weapon Assignment Algorithms for a SDI Mid-Course Battle Manager},
year={1990}
}

@ARTICLE{Murphey2000,
author={Murphey, R.A.},
title={An Approximate Algorithm For A Weapon Target Assignment Stochastic Program},
journal={Approximation and Complexity in Numerical Optimization: Continuous and Discrete Problems},
year={2000},
pages={406-421}
}

@CONFERENCE{Shang2008221,
author={Shang, G.},
title={Solving weapon-target assignment problems by a new ant colony algorithm},
journal={Proceedings of the 2008 International Symposium on Computational Intelligence and Design, ISCID 2008},
year={2008},
volume={1},
pages={221-224},
doi={10.1109/ISCID.2008.28}
}

@ARTICLE{Soland1973,
author={Soland, Richard M.},
title={Optimal Defensive Missile Allocation: A Discrete Min-Max Problem.},
journal={Operations Research},
year={1973},
volume={21},
number={2},
pages={590-596},
doi={10.1287/opre.6.3.346}
}

@ARTICLE{Tokgöz201339,
author={Tokgöz, A. and Bulkan, S.},
title={Weapon target assignment with combinatorial optimization techniques},
journal={International Journal of Advanced Research in Artificial Intelligence},
year={2013},
volume={2},
number={7},
pages={39-50}
}

@ARTICLE{Xin2011598,
author={Xin, B. and Chen, J. and Peng, Z. and Dou, L. and Zhang, J.},
title={An efficient rule-based constructive heuristic to solve dynamic weapon-target assignment problem},
journal={IEEE Transactions on Systems, Man, and Cybernetics Part A:Systems and Humans},
year={2011},
volume={41},
number={3},
pages={598-606},
doi={10.1109/TSMCA.2010.2089511}
}

@article{timing_cost_function,
author = {Volle, Kyle and Rogers, Jonathan},
title = {Weapon–Target Assignment Algorithm for Simultaneous and Sequenced Arrival},
journal = {Journal of Guidance, Control, and Dynamics},
volume = {41},
number = {11},
pages = {2361-2373},
year = {2018},
doi = {10.2514/1.G003515}
}

@ARTICLE{grouping_constraint,
  author={Guo, Dong and Liang, Zixuan and Jiang, Peng and Dong, Xiwang and Li, Qingdong and Ren, Zhang},
  journal={IEEE Access}, 
  title={Weapon-Target Assignment for Multi-to-Multi Interception With Grouping Constraint}, 
  year={2019},
  volume={7},
  pages={34838-34849},
  doi={10.1109/ACCESS.2019.2898874}
}

@ARTICLE{ucav_routing_problem,
  author={Jiaming, Zhang and Zhong, Liu and Jianmai, Shi and Chao, Chen},
  journal={Journal of Systems Engineering and Electronics}, 
  title={Weapon configuration, allocation and route planning with time windows for multiple unmanned combat air vehicles}, 
  year={2018},
  volume={29},
  number={5},
  pages={953-968},
  doi={10.21629/JSEE.2018.05.08}
}

@article{artillery_attack,
title = {Fire scheduling for planned artillery attack operations under time-dependent destruction probabilities},
journal = {Omega},
volume = {38},
number = {5},
pages = {383-392},
year = {2010},
issn = {0305-0483},
doi = {https://doi.org/10.1016/j.omega.2009.10.003},
author = {Young-Ho Cha and Yeong-Dae Kim}
}

@article{BidGroup,
author = {Kim, Keum-Seong and Kim, Ho-Yeon and Choi, Han-Lim},
journal = {International journal of aeronautical and space sciences},
number = {1},
pages = {290-302},
publisher = {The Korean Society for Aeronautical & Space Sciences (KSAS)},
title = {A Bid-Based Grouping Method for Communication-Efficient Decentralized Multi-UAV Task Allocation},
volume = {21},
year = {2020}
}

@inproceedings{RobotTaskAllocation,
author = {Sarkar, Chayan and Paul, Himadri Sekhar and Pal, Arindam},
booktitle = {IEEE International Conference on Robotics and Automation (ICRA)},
pages = {5022-5027},
title = {A Scalable Multi-Robot Task Allocation Algorithm},
year = {2018}
}

@article{MarketAlg,
author = {Nelke, Sofia Amador and Okamoto, Steven and Zivan, Roie},
journal = {ACM transactions on intelligent systems and technology},
number = {1},
pages = {1-25},
title = {Market clearing–based dynamic multi-agent task allocation},
volume = {11},
year = {2020}
}

@INPROCEEDINGS{One-to-many,
  author={Dutta, Ayan and Asaithambi, Asai},
  booktitle={2019 International Conference on Robotics and Automation (ICRA)}, 
  title={One-to-many bipartite matching based coalition formation for multi-robot task allocation}, 
  year={2019},
  pages={2181-2187},
  doi={10.1109/ICRA.2019.8793855}
}

@conference{Paradis2005,
	author = {Paradis, Stéphane and Benaskeur, Abderrezak and Oxenham, Martin and Cutler, Philip},
	title = {Threat evaluation and weapons allocation in network-centric warfare},
	year = {2005},
	journal = {International Conference on Information Fusion, FUSION},
	volume = {2},
	pages = {1078 – 1085},
	doi = {10.1109/ICIF.2005.1591977}
}

@article{Song2009,
	author = {Song, Zhihua and Zhu, Fashun and Zhang, Duolin},
	title = {A heuristic genetic algorithm for solving constrained weapon-target assignment problem},
	year = {2009},
	journal = {Proceedings - 2009 IEEE International Conference on Intelligent Computing and Intelligent Systems, ICIS 2009},
	volume = {1},
	pages = {336 – 341},
	doi = {10.1109/ICICISYS.2009.5357831}
}

@inproceedings{lloyd1986,
  title={Weapons allocation is NP-complete.},
  author={Lloyd, Stuart P and Witsenhausen, Hans S},
  booktitle={1986 summer computer simulation conference},
  pages={1054--1058},
  year={1986}
}

@ARTICLE{Murphey1999406,
author={Murphey, R.A.},
title={An approximate algorithm for a weapon target assignment stochastic program},
journal={Approximation and Complexity in Numerical Optimization: Continuous and Discrete Problems},
year={1999},
pages={406-421}
}

@article{survey2024,
title = {A comprehensive survey of weapon target assignment problem: Model, algorithm, and application},
journal = {Engineering Applications of Artificial Intelligence},
volume = {137},
pages = {109212},
year = {2024},
author = {Jinrui Li and Guohua Wu and Ling Wang},
doi = {https://doi.org/10.1016/j.engappai.2024.109212}
}

@article{quantum_alg,
title = {A quantum algorithm for solving weapon target assignment problem},
journal = {Engineering Applications of Artificial Intelligence},
volume = {125},
pages = {106668},
year = {2023},
author = {Erdi Acar and Saim Hatipoğlu and İhsan Yılmaz},
doi = {https://doi.org/10.1016/j.engappai.2023.106668}
}

@article{survey2025,
title = {Current development and future prospects of multi-target assignment problem: A bibliometric analysis review},
journal = {Defence Technology},
volume = {43},
pages = {44-59},
year = {2025},
author = {Shuangxi Liu and Zehuai Lin and Wei Huang and Binbin Yan},
doi = {https://doi.org/10.1016/j.dt.2024.09.006}
}

@article{uav_defence_technology,
title = {Weapon-target assignment for unmanned aerial vehicles: A multi-strategy threshold public goods game approach},
journal = {Defence Technology},
year = {2025},
issn = {2214-9147},
doi = {https://doi.org/10.1016/j.dt.2025.01.014},
author = {Wenhao Bi and Zhaoxi Wang and Yang Xu and An Zhang}
}

@article{Sam_Matlin1970,
 author = {Samuel Matlin},
 journal = {Operations Research},
 number = {2},
 pages = {334--373},
 publisher = {INFORMS},
 title = {A Review of the Literature on the Missile-Allocation Problem},
 volume = {18},
 year = {1970}
}

\end{document}